\documentclass[conference, final]{IEEEtran}

\usepackage[utf8]{inputenc}
\usepackage[T1]{fontenc}
\usepackage{microtype}
\usepackage[ngerman,english]{babel}
\usepackage[obeyFinal]{todonotes} 
\usepackage{listings} 
\usepackage{amsmath}
\usepackage[hidelinks]{hyperref}
\usepackage{cleveref}
\usepackage{multicol}
\usepackage{glossaries}
\usepackage[commandnameprefix=always,todonotes={textsize=tiny}]{changes}
\usepackage{tikz}
\usepackage[binary-units=true]{siunitx}
\usepackage{dblfloatfix}
\usepackage{physics}
\usepackage{csquotes}

\usetikzlibrary{calc,backgrounds,fit}

\usepackage[backend=biber,maxbibnames=1,minbibnames=1,mincitenames=1,giveninits=true,sorting=none,style=ieee,doi=true,eprint=false,isbn=false,url=false]{biblatex}
\newacronym{adc}{ADC}{analog-to-digital converter}
\newacronym{adex}{AdEx}{adaptive exponential leaky integrate-and-fire}
\newacronym{ann}{ANN}{artificial neural network}
\newacronym{asic}{ASIC}{application-specific integrated circuit}
\newacronym{asicab}{\acrshort{asic} adapter \acrshort{pcb}}{\acrlong{asic} adapter \acrlong{pcb}}
\newacronym{api}{API}{application programming interface}
\newacronym{bmbf}{BMBF}{German Federal Ministry of Education and Research}
\newacronym{bss2}{\mbox{BSS-2}}{Brain\mbox{ScaleS-2}}
\newacronym{bss1}{\mbox{BSS-1}}{Brain\mbox{ScaleS-1}}
\newacronym{bss2os}{\gls{bss2} OS}{\gls{bss2} Operating System}
\newacronym{bss}{BSS}{BrainScaleS}
\newacronym{cpu}{CPU}{central processing unit}
\newacronym{dfki}{DFKI}{German Research Centre for Artificial Intelligence}
\newacronym{dma}{DMA}{direct memory access}
\newacronym{dram}{DRAM}{dynamic random-access memory}
\newacronym{fpga}{FPGA}{field-programmable gate array}
\newacronym{isa}{ISA}{instruction set architecture}
\newacronym{jit}{JIT}{just-in-time}
\newacronym{lif}{LIF}{leaky integrate-and-fire}
\newacronym{madc}{MADC}{membrane analog-to-digital converter}
\newacronym{mse}{MSE}{mean squared error}
\newacronym{cadc}{CADC}{columnar analog-to-digital converter}
\newacronym{pcb}{PCB}{printed circuit board}
\newacronym{ppu}{PPU}{plasticity processing unit}
\newacronym{simd}{SIMD}{single instruction, multiple data}
\newacronym{snn}{SNN}{spiking neural network}
\newacronym{sram}{SRAM}{static random-access memory}
\newacronym{stdp}{STDP}{spike timing dependent plasticity}
\newacronym{stp}{STP}{short term plasticity}
\newacronym{nasprop}{NASProp}{neuromorphic accumulative spike propagation}
\newacronym{vu}{VU}{vector unit}
\newacronym{stc}{STC}{synaptic tagging-and-capture}
\newacronym{ltm}{LTM}{long-term memory}
\newacronym{prp}{PRP}{plasticity-related protein}
\newacronym{ltp}{LTP}{long-term potentiation}
\newacronym{ltd}{LTD}{long-term depression}
\newacronym{sr}{SR}{stochastic rounding}
\newacronym{su}{SU}{stochastic updates}
\newacronym{prng}{PRNG}{pseudorandom number generator}
\newacronym{stet}{STET}{strong tetanic stimulation}
\newacronym{wtet}{WTET}{weak tetanic stimulation}
\newacronym{slfs}{SLFS}{strong low-frequency stimulation}
\newacronym{wlfs}{WLFS}{weak low-frequency stimulation}

\newcommand\copyrighttext{%
       \footnotesize This article has been accepted at the \textit{Neuro Inspired Computational Elements Conference 2025} and will appear in the conference proceedings.\\
       \textcopyright 2025 IEEE. Personal use of this material is permitted.
       Permission from IEEE must be obtained for all other uses, in any current or future media, including reprinting/republishing this material for advertising or promotional purposes, creating new collective works, for resale or redistribution to servers or lists, or reuse of any copyrighted component of this work in other works.}
\newcommand\copyrightnotice{%
    \begin{tikzpicture}[remember picture,overlay]
        \node[anchor=south,yshift=10pt] at (current page.south) {\fbox{\parbox{\dimexpr\textwidth-\fboxsep-\fboxrule\relax}{\copyrighttext}}};
    \end{tikzpicture}%
}

\begin{document}

\title{Multi-timescale synaptic plasticity on\\analog neuromorphic hardware}

\author{%
	\IEEEauthorblockN{%
		Amani Atoui\IEEEauthorrefmark{1}\IEEEauthorrefmark{3}, Jakob Kaiser\IEEEauthorrefmark{1}, Sebastian Billaudelle\IEEEauthorrefmark{2}\IEEEauthorrefmark{1}, Philipp Spilger\IEEEauthorrefmark{1}, Eric Müller\IEEEauthorrefmark{1},\\
		Jannik Luboeinski\IEEEauthorrefmark{4}, Christian Tetzlaff\IEEEauthorrefmark{4} and Johannes Schemmel\IEEEauthorrefmark{1}}%
	\IEEEauthorblockA{\IEEEauthorrefmark{1}\textit{Kirchhoff Institute for Physics}, Heidelberg University, Germany}%
    \IEEEauthorblockA{\IEEEauthorrefmark{2}\textit{Institute for Neuroinformatics}, University of Zurich and ETH Zürich, Zürich, Switzerland}%
    \IEEEauthorblockA{\IEEEauthorrefmark{4}\textit{Group of Computational Synaptic Physiology}, Department of Neuro- and Sensory Physiology,\\ University Medical Center Göttingen, Germany}%
	\IEEEauthorblockA{\IEEEauthorrefmark{3}\href{mailto:amani.atoui@kip.uni-heidelberg.de}{amani.atoui@kip.uni-heidelberg.de}}%

}

\maketitle
\copyrightnotice
\begin{abstract}
As numerical simulations grow in complexity, their demands on computing time and energy increase.
Accelerators for numerical computation offer significant efficiency gains in many computationally-intensive scientific fields, but their use in simulating spiking neural networks in computational neuroscience is hindered by challenges, mainly in effective parallelism and efficient use of memory in the presence of sparse representations and sparse communication.
The BrainScaleS architectures are neuromorphic substrates that can emulate spiking neural networks at accelerated timescales compared to real time, which offers an advantage for studying complex plasticity rules that require extended simulation runtimes.
This work presents the implementation of a calcium-based plasticity rule that integrates calcium dynamics based on the synaptic tagging-and-capture hypothesis on the BrainScaleS-2 system.
The implementation of the plasticity rule for a single synapse involves incorporating the calcium dynamics and the plasticity rule equations.
The calcium dynamics are mapped to the analog circuits of BrainScaleS-2, while the plasticity rule equations are numerically solved on its embedded digital processors.
The main hardware constraints include the speed of the processors and the use of integer arithmetic.
By adjusting the timestep of the numerical solver and introducing stochastic rounding, we demonstrate that \mbox{BrainScaleS-2} accurately emulates a single synapse following a calcium-based plasticity rule across four established stimulation protocols and validate our implementation against a software reference model.

\end{abstract}


\section{Introduction}\label{sec:introduction}
Computer simulations are an indispensable part of computational neuroscience, specifically for understanding neuron circuits through synaptic plasticity \cite{zenke2014auryn}.
The challenge lies in simulating complex models and large networks with an adequate simulation timestep for the numerical solver to capture milliseconds neuronal activity.
This can result in long waiting times when simulating long-term synaptic plasticity which operates in the timescale of several hours \cite{zenke2014auryn}.
There comes the role of specialized neuromorphic hardware, such as \gls{bss2}, that can emulate neural dynamics at accelerated timescales compared to real time.

Our aim is to employ the advantages of neuromorphic hardware for the understanding of synaptic plasticity through calcium-based models.
Molecular studies have identified calcium as a key element of plasticity mechanisms, which has motivated its integration into mechanistic models of long-term synaptic plasticity \cite{graupner2012calcium,li2016induction}.
Such models can describe a diversity of plasticity curves that have been observed experimentally in different systems \cite{graupner2012calcium,sajikumar2005synaptic}, as well as provide insights into memory consolidation in recurrent neural networks \cite{luboeinski2024modeling}.

The \gls{stc} hypothesis provides an explanation of the neuronal and synaptic processes underlying memory consolidation based on evidence from molecular studies \cite{redondo2011making, shires2012synaptic, frey1997synaptic}.
This hypothesis has also been linked to so-called behavioral tagging experiments \cite{moncada2007induction,moncada2015behavioral}, relating the \gls{stc} hypothesis to the study of learning and memory in cognitive neuroscience.
Theoretical studies of single synapses \cite{clopath2008tag}, feedforward neural networks \cite{ziegler2015synaptic}, and recurrent neural networks \cite{luboeinski2021memory} have supported the \gls{stc} hypothesis and its role in memory consolidation (see \cite{luboeinski2024modeling} for a review).

In this work, we implement a plasticity rule that obeys the \gls{stc} hypothesis for a single synapse on neuromorphic hardware.
Based on this, our end goal is to provide an accelerated implementation of spiking neural networks whose plasticity rule follows the \gls{stc} hypothesis to answer scientific questions about learning and memory.
We start by giving an overview of the \gls{bss2} system and the plasticity model \cite{luboeinski2021memory} that we adopt. We also describe useful concepts for our implementation such as the \gls{sr} scheme.
We then list the hardware constraints that we encounter in the implementation and our propositions to overcome them.
Our key contributions include updating different model variables on multiple timescales and using \gls{sr} for integer arithmetic.
The results show that the considered plasticity rule can be successfully emulated for a single synapse on \gls{bss2} due to the compensation across the effects that arise from analog computing and stochasticity across spike trials.

\subsection{The \acrlong{bss2} System}
On the neuromorphic \gls{bss2} system, neuron and synapse dynamics are emulated by analog circuits.
The \num{512} neuron circuits on the system replicate the behavior of the \gls{adex} neuron model~\cite{brette2005adaptive} in continuous time.
\Gls{bss2} emulates these dynamics at a tunable acceleration factor of \num{1000} compared to biological real time \cite{billaudelle2022accurate}.
The \gls{adex} model consists of two coupled differential equations for the membrane potential $V(t)$ and the adaptation current $I_\text{adapt}(t)$. The membrane potential is given by
\begin{equation}
	\label{eq:adex}
	\begin{split}
		C_\text{m} \dv{V(t)}{t} =&\ g_\text{L} \cdot \left( V_\text{L} - V(t) \right) \\
		                        & + g_\text{L} \Delta_\text{T} \exp\left( \frac{V(t) - V_\text{T}}{\Delta_\text{T}} \right) \\
								& + I_\text{syn}(t) - I_\text{adapt}(t),
	\end{split}
\end{equation}
where $C_\text{m}$ is the membrane capacitance, $g_\text{L}$ the leak conductance, $V_\text{L}$ the leak potential, $\Delta_\text{T}$ the threshold slope factor and $V_\text{T}$ the effective threshold potential.

The second equation describes the dynamics of the adaptation current
\begin{equation}
	\label{eq:adapt}
		\tau_\text{adapt} \dv{I_\text{adapt}(t)}{t} = a \left( V_\text{m}(t) - V_\text{L} \right) - I_\text{adapt}(t),
\end{equation}
where $\tau_\text{adapt}$ is the adaptation time constant and $a$ the subthreshold adaptation.

$I_\text{syn}(t)$ represents the synaptic input current.
\Gls{bss2} supports conductance- as well as current-based synapses \cite{billaudelle2022accurate}.
In the presented work, we will use current-based synapses, which implement an exponentially decaying current for each input spike, as in \cite{luboeinski2021memory}.

Once the membrane potential $V$ reaches a threshold potential $V_\text{th}$ the membrane potential is clamped to the reset potential $V_\text{reset}$ for the duration of the refractory period $\tau_\text{ref}$ and the adaptation current $I_\text{adapt}$ is increased by the spike-triggered adaptation $b$: $I_\text{adapt} \rightarrow I_\text{adapt} + b$.
Furthermore, a digital spike event is generated and forwarded to postsynaptic partners.
Each neuron circuit is connected to \num{256} synapse circuits storing \SI{6}{\bit} weights which convert the digital, presynaptic events to voltage pulses which control the synaptic current on each membrane.
Neurons can also be configured such that each presynaptic spike causes a postsynaptic spike, the so-called ``bypass'' mode.

The modular design of the neuron circuits allows to selectively enable and disable features of the neuron.
For example, the exponential and adaptation terms can be disabled to reduce the neuron model to a \gls{lif} neuron model.
In addition, an analog dynamic memory array \cite{hock13analogmemory} allows to adjust each of the parameters mentioned in \cref{eq:adex,eq:adapt} individually for each neuron.

\begin{figure}
	\includegraphics[width=\columnwidth]{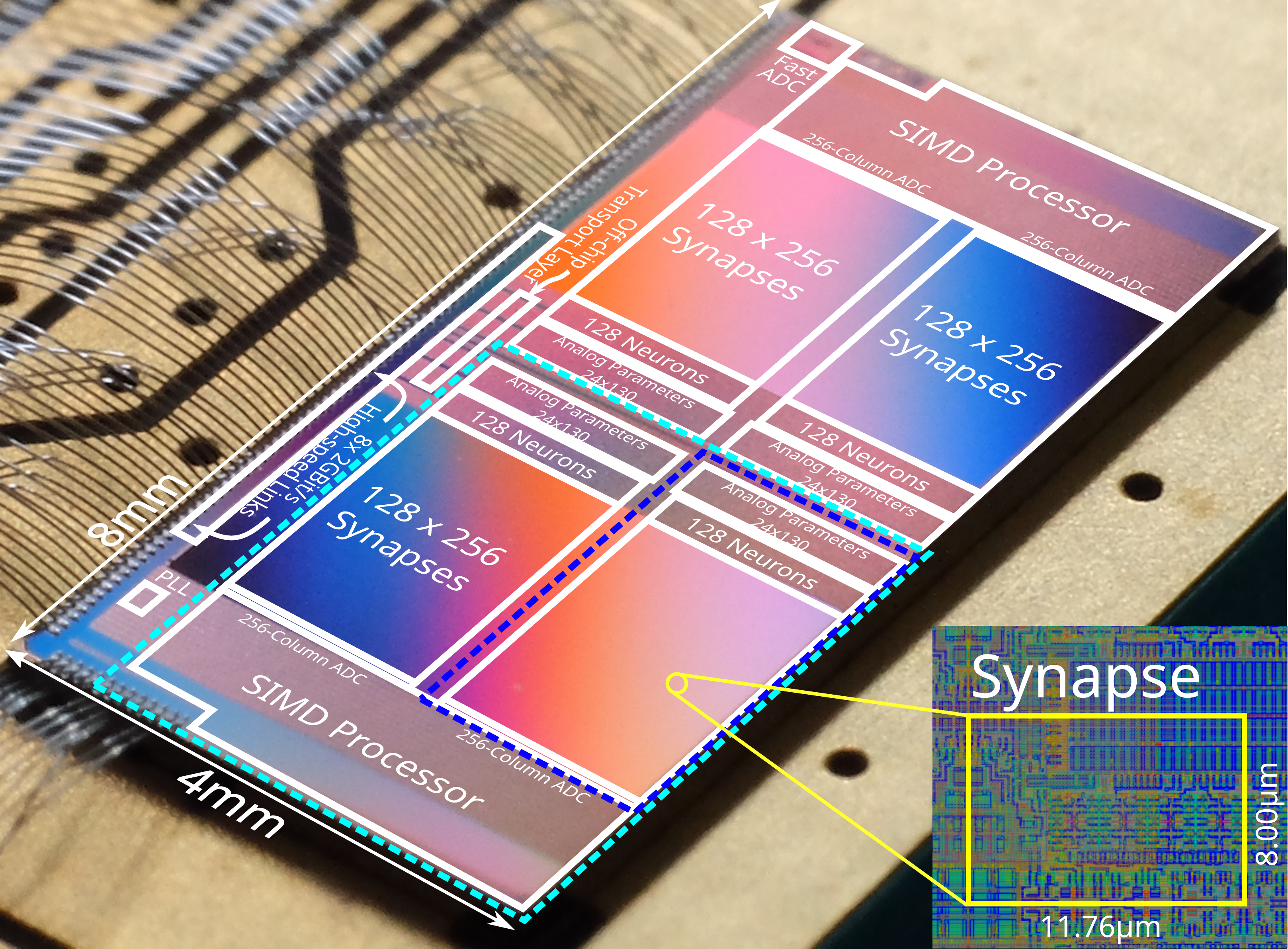}
	\caption{%
		\label{fig:chipphoto}
		Schematic and chip photo of the \gls{bss2} \acrfull{asic}.
		One embedded \gls{simd} processor per chip half can run arbitrary code to access on-chip observables, such as membrane potentials, and to perform changes to topology, neuron and synapse parameterization.
	}
\end{figure}

The \gls{bss2} \gls{asic} is organized in two halves with \num{256} neurons at the top and \num{256} neurons at the bottom, see \cref{fig:chipphoto}.
Each half of the chip also hosts a digital processor.
These so-called \glspl{ppu} implement the \mbox{Power} \gls{isa} \mbox{2.06}~\cite{powerisa_206} for \SI{32}{\bit} and a custom vector unit extension.
The \glspl{ppu} can access a vector-parallel \gls{cadc}, which enables to sample, for example, the adaptation currents of several neurons in parallel.
Furthermore, it can modify the neuron parameters as well as the synaptic weights~\cite{pehle2022brainscales2_nopreprint_nourl}.
While the \glspl{ppu} do not provide direct access to spikes and \acrfull{madc} data, the data can in principle be recorded to \gls{fpga}-based \gls{dram} which is accessible to the embedded processors and the host computer.

The \gls{bss2} software stack provides a high-level modeling interface for experiment definition and a \mbox{C++}~\cite{mueller2022scalable_noeprint} runtime with a focus on embedded real-time programming to express programmable processes and plasticity rules on the \gls{ppu}.

\subsection{Memory and Synaptic Tagging-and-Capture Hypothesis}
An important physical substrate for the emergence of \gls{ltm} is the stabilization of synaptic efficacies within neural networks, referred to as synaptic consolidation.
The biological mechanisms proposed to underlie synaptic consolidation involve two different phases \cite{redondo2011making,okuda2021initial,prodan2024synaptic}:
the early phase is characterized by receptor alterations and turnover based on the calcium concentration in the postsynapse, while the late phase is characterized by the addition of receptor slots in the case of potentiation and their removal in the case of depression.
The transfer from the early-phase to the late-phase is achieved by biophysical and structural alterations in the synapses that contribute to stabilizing the levels of synaptic strength.

The \gls{stc} hypothesis \cite{frey1997synaptic, redondo2011making} is a specific theory that has been found to explain main aspects of synaptic consolidation \cite{okuda2021initial,prodan2024synaptic,luboeinski2024modeling}.
It adopts the concept that early-phase plasticity creates a potential for stable synaptic weight changes while not being stable itself \cite{redondo2011making,luboeinski2021memory}.
According to the \gls{stc} hypothesis, late-phase plasticity occurs after neurons have fired sufficiently. It consists of the following steps.
First, early-phase plasticity is expressed, and the synapse is marked with a local synaptic tag for specificity.
Second, proteins are synthesized and distributed.
Third, these proteins are captured by the tagged synapses which allow the stabilization of the synaptic strength in the late phase.

The time course of synaptic consolidation is influenced by the recent history and near future of the network activity \cite{frey1997synaptic, redondo2011making}, thus, it involves a set of mechanisms that can but do not have to be triggered at a single moment in time. 
Through the interaction of the network with the synthesized proteins and synaptic tags, the \gls{stc} hypothesis accounts for the time window of inital (synaptic) memory consolidation.

\subsection{Plasticity Equations}
We use the synapse model from \cite{luboeinski2021memory} that integrates the calcium dynamics with the \gls{stc} hypothesis.
The neuron dynamics follow the \gls{lif} model. For our work, we consider a single synapse between a presynaptic neuron $j$ and a postsynaptic neuron $i$.
The calcium dynamics of the postsynaptic site are given by
\begin{align}
\label{calcium-eqn}
    \dv{c(t)}{t}= &-\frac{{c}(t)}{{\tau }_{{\rm{c}}}}+{c}_{{\rm{pre}}}\ \mathop{\sum}\limits _{n}\delta (t-{t}_{j}^{n}-{t}_{{\rm{c,delay}}})\\&+{c}_{{\rm{post}}}\ \mathop{\sum}\limits _{m}\delta (t-{t}_{i}^{m}), \nonumber
\end{align}
with the calcium time constant ${{\tau }_{{\rm{c}}}}$, the presynaptic spike times $t_j^n$, the postsynaptic spike times $t_i^m$, the contribution of presynaptic spikes ${c}_{{\rm{pre}}}$, the contribution of postsynaptic spikes ${c}_{{\rm{post}}}$, and the delay of calcium concentration triggered by presynaptic spikes ${t}_{{\rm{c,delay}}}$.

The dynamics of the early-phase weight $h(t)$ of the synapse (given in units of nanocoulomb) are governed by
\begin{align}
\label{early-phase-eqn}
    {\tau }_{h}\dv{h(t)}{t} = &\;0.1\ ({h}_{0}-{h}(t))\\&+{\gamma }_{{\rm{p}}}(h_{\rm{max}}-{h}(t))\cdot {{\Theta }}[{c}(t)-{\theta }_{{\rm{p}}}]\nonumber\\ &-{\gamma }_{{\rm{d}}}({h}(t)-h_{\rm{min}})\cdot {{\Theta }}[{c}(t)-{\theta }_{{\rm{d}}}]+\xi (t), \nonumber
\end{align}
with $\Theta[\cdot]$ being the Heaviside function and ${\tau }_{h}$ being a time constant.
The first term of \cref{early-phase-eqn} describes a relaxation of the early-phase weight to its initial value ${h}_{0}$, the second term describes early-phase \gls{ltp} with a rate ${\gamma }_{{\rm{p}}}$ for calcium concentration above the potentiation threshold ${\theta }_{{\rm{p}}}$, and the third term describes early-phase \gls{ltd} with a rate ${\gamma }_{{\rm{d}}}$ for calcium concentration above the depression threshold ${\theta }_{{\rm{d}}}$.
The term $\xi (t)$ describes the calcium-dependent noise-driven fluctuations.
The weight dynamics are bounded by $h_{\rm{max}}=\SI{1}{\nano\coulomb}$ and $h_{\rm{min}}=\SI{0}{\nano\coulomb}$.

The protein amount $p(t)$ for the postsynaptic neuron is updated using
\begin{equation}
    \label{protein-eqn}
    {\tau }_{p}\dv{p(t)}{t}=-{p}(t)+\alpha \ {{\Theta }}\left[\left|{h}(t)-{h}_{0}\right|-{\theta }_{{\rm{pro}}}\right],
\end{equation}
where $\alpha$ is the protein synthesis rate and ${\theta }_{{\rm{pro}}}$ is the protein synthesis threshold.

The dynamics of the late-phase weight $z(t)$ of the synapse depend on the protein amount $p(t)$, early-phase weight $h(t)$, and a tagging threshold ${\theta }_{{\rm{tag}}}$ as
\begin{align}
    \label{late-phase-eqn}
    {\tau }_{z}\dv{z(t)}{t}=&\ {p}(t)\cdot (z_{\rm{max}}-{z}(t))\cdot {{\Theta }}[({h}(t)-{h}_{0})-{\theta }_{{\rm{tag}}}]\\&-{p}(t)\cdot ({z}(t)-z_{\rm{min}})\cdot{{\Theta }}[({h}_{0}-{h}(t))-{\theta }_{{\rm{tag}}}].\nonumber
\end{align}
The late-phase weight is bounded by $z_{\rm{max}}=1$ and $z_{\rm{min}}=-0.5$.
Finally, the total synaptic weight $w(t)$ is given by
\begin{equation}
    \label{total-weight-eqn}
    {w}(t)={h}(t)+{h}_{0}\cdot{z}(t).
\end{equation}

\subsection{Integer Arithmetic}\label{sr:section}
The \gls{ppu} does not feature hardware support for floating-point arithmetic~\cite{kahan1996ieee}.
Instead, numeric calculations have to be implemented using \SI{8}{bit} and \SI{16}{bit} integers.
This means that variables are represented using a maximum dynamic range of \mbox{$[0 \ldotp \ldotp \text{MAX\_UINT}_n= 2^n-1]$} for unsigned \textit{n}-bit integers and \mbox{$[\text{MIN\_INT}_n = -2^{n-1} \ldotp \ldotp \text{MAX\_INT}_n= 2^{n-1}-1]$} for signed \textit{n}-bit integers.
This reduced arithmetic is advantageous in terms of silicon space, energy, and memory storage, but it also imposes challenges in terms of resolution and truncation \cite{hopkins2020stochastic}.
By default, operations on integers implement the \textit{truncate-to-zero} mode, which becomes problematic when adding fractional amounts or dividing by integers; this mode typically accumulates errors and produces underestimated final results.

Other challenges can be a consequence of the programming language.
Because C and C++ focus on processing speed, they open up many possibilities for triggering undefined behavior when performing integer arithmetic operations~\cite{dietz2015understanding,wang2013towards}.

To overcome the numerical errors from using integer arithmetic, the stochastic rounding mode was suggested in \cite{forsythe1950round}.    
In the \gls{sr} mode, a given number is mapped to one of the two nearest representable numbers with a probability that depends on the distances of the given number to the nearest numbers.%
%

There are two main situations where \gls{sr} is useful.
First, \gls{sr} ensures zero-mean rounding errors and produces smaller errors especially in situations where \textit{truncate-to-zero} produces rounding errors of one sign \cite{croci2022stochastic}.
Suppose an integer $x$ is multiplied with a fraction $\frac{n}{d}$. The result with \textit{truncate-to-zero} is an integer $y$ such that
\begin{equation}
    y=\lfloor{x\cdot \frac{n}{d}}\rfloor,
\end{equation}
which accumulates an error
\begin{equation}
    \epsilon = \frac{(x\cdot n)\bmod d}{d}.
\end{equation}
This can be overcome by stochastically compensating for the error
\begin{equation}
    y'=
    \begin{cases}
        \lceil{x\cdot \frac{n}{d}}\rceil = y+1 \text{ with probability } \epsilon, \\
        \lfloor{x\cdot \frac{n}{d}}\rfloor = y \text{ with probability } 1-\epsilon.
    \end{cases}
    \label{sr-round-eqn}
\end{equation}
Second, \gls{sr} is immune to stagnation, a phenomenon where a sequence of small updates relative to large quantities is lost if these changes cannot be represented by the machine's number format.
Suppose an integer $x$ is successively updated by a fixed amount $\epsilon<1$ such that
\begin{equation}
    x'=x+\epsilon .
\end{equation}
Since $\epsilon$ cannot be represented by an integer, \gls{sr} proposes updating $x$ according to
\begin{equation}
    x'=
    \begin{cases}
        x+1 \text{ with probability } \epsilon, \\
        x \text{ with probability } 1-\epsilon.
    \end{cases}
    \label{sr-update-eqn}
\end{equation}

While \gls{sr} has been known since the 1950s, the interest in it is currently expanding as it proved useful in applications mainly related to machine learning and solving ordinary differential equations on reduced precision systems.
\gls{sr} was proposed to overcome the resulting numerical drawbacks and was proven useful in different use cases such as solving ordinary differential equations to produce accurate spike timings in the Izhikevich neuron model \cite{hopkins2020stochastic}.

\section{Methods}\label{sec:methods}
In this work, we exploit the concept of updates at different timesteps. Unlike numerical simulations on conventional hardware, which typically define a unified timestep to update all the model variables, we use different timesteps depending on the involved hardware and time constants.
This is done since part of the variables are emulated in continuous time in the analog core of \gls{bss2} while others are solved numerically on an embedded processor.
Specifically, the membrane potential and calcium concentration will be emulated as continuous-time analog signals, the transient early-phase weight will be updated regularly, and the early-phase weight relaxation, late-phase weight, and protein amount will be updated stochastically.
\begin{figure}
	\includegraphics[width=\columnwidth]{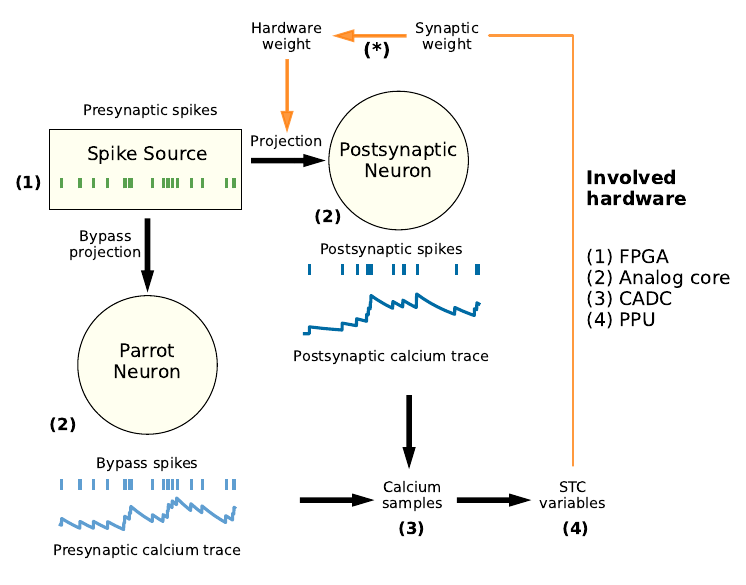}
	\caption{
        Experimental setup for emulating a single synapse following the considered plasticity rule on \gls{bss2}.
		The \gls{fpga} provides a real-time capable memory-buffered input and output interface for the neuromorphic chip.
		A spike source is placed on the \gls{fpga} and used to generate presynaptic spikes.
        These are forwarded to the postsynaptic neuron via a projection whose weight update obeys the considered plasticity rule.
		The same presynaptic spikes are mirrored to the parrot neuron to emulate the presynaptic calcium dynamics.
		The ``bypass'' mode of the parrot neuron ensures that presynaptic spikes are immediately translated to postsynaptic spikes.
        The weighted sum of the adaptation traces of the postsynaptic and parrot neurons represents the calcium trace.
        Calcium samples obtained by the \gls{cadc} are then used by the \gls{ppu} to calculate the model variables, including the synaptic weight, which will be used to update the weight of the projection.
        The mapping of the synaptic weight to hardware indicated by (*) has a negligible effect in the considered single synapse case, so we chose not to implement it in this work.
    }
    \label{fig:overview}
\end{figure}

\subsection{Experimental Setup on \acrlong{bss2}}
The experimental setup for the emulation of the plasticity rule is shown in \cref{fig:overview}. Here, we explain the intuition behind our approach and discuss the details of the involved hardware.

The neuron dynamics are emulated by the analog core of \gls{bss2} while the synaptic weights can be computed numerically on a fixed time grid using the \gls{ppu}.
An important consideration here is to link the fast timescale of the neuron dynamics and spikes to the slow timescale of the synaptic weight dynamics.
This link is achieved by the calcium concentration that directly depends on the spikes, but also drives the updates of the synaptic weights.
Therefore, we map the calcium concentration to the analog circuit using the adaptation current (\cref{eq:adapt}) whose dynamics greatly resemble those of the calcium concentration (\cref{calcium-eqn}).
The calcium traces are then sampled from the analog core at a fixed sampling period \textit{$\Delta t$}, and these samples are used to update the synaptic weights.
In this way, the calcium concentration acts as a domain crossing that maps the fast neuron dynamics to the slow synaptic weight dynamics.



To set up the experiment on \gls{bss2}, a spike source is configured that plays the role of the presynaptic neuron which injects spikes into a postsynaptic neuron.
For the considered plasticity rule, additional requirements apply.
The adaptation trace generated by the postsynaptic neuron depends on the postsynaptic spikes only, but the calcium trace depends on the presynaptic and postsynaptic spikes.
To track the contribution of presynaptic spikes, we exploit the linearity of \cref{calcium-eqn} which allows us to split the equation into a presynaptic component and a post-synaptic component.
We use an additional neuron, called \enquote{parrot neuron}, to account for the contribution of the presynaptic spikes.
This circuit is connected to the same spike source and is configured in the ``bypass'' mode such that every input spike triggers an output spike.
In this way, we mirror the presynaptic spikes and obtain an adaptation trace from the spikes of the parrot neuron.
The required calcium trace is thus the weighted sum of the two adaptation traces.

Simulation results showed that the calcium contribution of postsynaptic spikes is negligible compared to that of the presynaptic spikes in the given stimulation protocols.
Thus, the main focus of this work is the implementation of the calcium dynamics in general as well as the plasticity equations.
The synaptic weight is not mapped to hardware, but is set to a small value such that no post-synaptic spikes occur.

\subsection{Stimulation Protocols}
To model the different forms of plasticity that could occur in a single synapse, we use the standard stimulation protocols that are generally adopted in the literature (provided in the supplementary information of \cite{luboeinski2021memory}).
Tetanic (high-frequency) stimulations induce \gls{ltp} due to the high calcium concentration released at the synapse while low-frequency stimulations induce \gls{ltd} due to the moderate calcium concentration.
Furthermore, strong stimulations cause the synapse to be marked by a local synaptic tag and the synthesis of a sufficient amount of \glspl{prp} necessary for late-phase synaptic plasticity whereas weak stimulations only induce synaptic tagging and early-phase plasticity. 
Consequently, the \gls{stet} protocol induces early-phase and late-phase \gls{ltp} whereas \gls{wtet} protocol induces only early-phase \gls{ltp}.
On the other hand, the \gls{slfs} protocol induces early-phase and late-phase \gls{ltd} whereas the \gls{wlfs} protocol induces only early-phase \gls{ltd}.

\subsection{Hardware Constraints and Limitations}
Using \gls{bss2} for implementing the plasticity rule imposes constraints that we should resolve:
\begin{enumerate}
	\item Plasticity rule timestep: this timestep accounts for the time required by the \gls{ppu} to sample from the analog traces and compute the weight update.
	The timestep of the numerical solver used in the model simulation in \cite{luboeinski2021memory} was \SI{0.2}{\milli\second} which is equivalent to \SI{0.2}{\micro\second} in our experiment parameteriziation on \gls{bss2}.
	This interval is too short for the current implementation on the \gls{ppu}, so we aimed to use slower updates.
    On the other hand, performing the numerical integration of the plasticity equations at long timesteps does not capture the required dynamics and yields distorted results.
    To ensure the expected behavior of the stimulation protocols in terms of early and late \gls{ltp} and \gls{ltd}, we found that this timestep must not exceed \SI{100}{\micro\second} and preferably remain below \SI{50}{\micro\second}.
    \item Integer arithmetic: The simulations in \cite{luboeinski2021memory} were performed using double-precision floating-point arithmetic, which is not supported by the \gls{ppu}.
	While floating-point numerics can be simulated in software, we prioritize using native integer arithmetic for speed purposes.
	Specifically, we use \SI{8}{\bit} integers for representing the model variables, and \SI{32}{\bit} integer arithmetic for operations and random number generation.
	This necessitates converting the model variables to a suitable dynamic range and further approximating model parameters, but also imposes truncation errors.
	\item Large time constants: these lead to slow changes of the model variables such as the protein amount, the late-phase weight, and the steady-state term of the early-phase weight.
	This leads to stagnation since the changes in the state variables and synaptic weights are small and cannot be represented by \SI{8}{\bit} integers.
	To resolve this, we suggest using a stochastic-update method.
\end{enumerate}
Besides these constraints, there remain distortions that are not resolved for the current emulation of the neuron and calcium dynamics:
\begin{enumerate}
	\item The postsynaptic calcium influx delay arising from the presynaptic spikes, described in \cref{calcium-eqn} as \textit{$t_{c,delay}$} will not be implemented.
	The reason is that the calcium traces are generated by the adaptation traces whose circuits are designed to increment the adaptation term at the spike time with no possibility for a predefined delay.
	\item The axonal spike delay will not be implemented as the circuits for emulating neuron dynamics do not account for this delay.
	\item The noise fluctuations of calcium during potentiation and depression that affect the early-phase weight, referred to as \textit{$\xi(t)$} described in \cref{early-phase-eqn}, will not be applied since it is computationally expensive.
\end{enumerate}

As we will show below, however, these limitations do not substantially affect our emulation of early- and late-phase synaptic plasticity dynamics on \gls{bss2}.

\subsection{Mapping Calcium to Adaptation Current}
As mentioned earlier, we split the calcium concentration into pre- and postsynaptic components and use the adaptation term to emulate the two components.
What makes the use of the adaptation term possible is that it can be enabled without affecting the neuron dynamics.
For each component, we set the subthreshold adaptation strength $a=0$ since the calcium concentration does not depend on the membrane potential.
The time constant \textit{$\tau_\text{adapt}$} is mapped to the calcium time constant \textit{$\tau_{c}$}.
The spike-triggered increment \textit{$b$} should be mapped to the contributions of presynaptic and postsynaptic spikes \textit{$c_\text{pre}$} and \textit{$c_\text{post}$}.
An important consideration is that \textit{$b$} should also be tuned to ensure a high dynamic range with a linear behavior of the circuit.

For tuning the adaptation parameters and comparing against the biological calcium trace in \cref{calcium-eqn}, we rely on the \gls{madc}.
Although the \gls{madc} will not be used for sampling the calcium traces when running the plasticity rule on \gls{bss2}, it has a higher sampling rate and a finer resolution compared to the \gls{cadc}.
Therefore, the adaptation traces can be retrieved, and the parameters can be fine-tuned to match the theoretical calcium trace.
We use the experimental setup described in \cref{fig:overview} and a stimulation scheme of Poisson spikes at \SI{100}{\kHz} for the tuning.
While running the plasticity rule, the calcium traces will be sampled using the \gls{cadc} at a lower sampling rate.

\subsection{Mapping Biological Parameters to Hardware}
For the emulation on \gls{bss2}, we map all parameters in the plasticity equations to hardware.
Parameters concerned with time dynamics such as $\tau_h$, $\tau_p$, $\tau_z$, and $\tau_c$ are divided by the acceleration factor of \num{1000}, while multiplicative factors such as $\gamma_p$ and $\gamma_d$ are kept constant.
As mentioned previously, the model variables are represented using integer arithmetic.
Specifically, the early-phase weight $h$ and protein amount $p$ are represented using unsigned \SI{8}{\bit} integers, while the late-phase weight $z$ is represented using a signed \SI{8}{\bit} integer.
The initial or steady-state values $h_0=\SI{0.420075}{\nano\coulomb}$, $ h_{\rm{max}}=\SI{1}{\nano\coulomb}$, $h_{\rm{min}}=\SI{0}{\nano\coulomb}$, $z_{\rm{min}}=-0.5$, and $z_{\rm{max}}=1$ are mapped to the ranges of their corresponding variables.
This yields $h_0^{\rm{hw}}=\lfloor{h_0\cdot \text{MAX\_UINT}_{8}}\rfloor$, $h_{\rm{max}}^{\rm{hw}}= \text{MAX\_UINT}_{8}$, $h_{\rm{min}}^{\rm{hw}}= \text{MIN\_UINT}_{8}$, $z_{\rm{min}}=0.5\cdot \text{MIN\_INT}_8$, and $z_{\rm{max}}=\text{MAX\_UINT}_{8}$.
Similarly, the protein and tagging thresholds $\theta_{\rm{pro}}$ and $\theta_{\rm{tag}}$ are mapped to the range of $h$.
Since we tune the calcium dynamics on the adaptation circuit, the calcium thresholds $\theta_p$ and $\theta_d$ are mapped experimentally using linear regression. 
\subsection{Solving the Differential Equations}
To solve the differential equations of \textit{h}, \textit{p}, and \textit{z} in the plasticity kernel, we use the explicit Euler method.
For a first-order differential equation $\dv{y(t)}{t} = F(t,y)$ and regular time step $\Delta t$, we can write the explicit Euler formula as
\begin{equation}
    y(t_n+\Delta t)=y(t_n )+ \Delta t \cdot F(t_n,y(t_n)).
	\label{Euler}
\end{equation}

To overcome the numerical errors from \textit{truncate-to-zero} in solving the differential equations, we apply \gls{sr}.
We use random \SI{32}{bit} unsigned integers drawn from a \SI{32}{bit} xorshift \gls{prng} \cite{marsaglia2003xorshift}.
The probabilities are then converted to integers by multiplying each probability with \mbox{MAX\_UINT$_{32}$} at runtime to obtain \textit{probability-equivalents}, and the \SI{32}{bit} random numbers are compared against these probability-equivalents.

\subsubsection{Early-phase weight}
Using \cref{Euler}, the differential equation for the early-phase weight \textit{h} in \cref{early-phase-eqn} can be rewritten as
\begin{align}
    {h}(t+\Delta t) &= h(t) + \frac{0.1\cdot\Delta t}{\tau_h}\ ({h}_{0}-{h}(t))\\ &+\frac{{\gamma }_{{\rm{p}}}\cdot\Delta t}{\tau_h}\cdot(255-{h}(t))\cdot {{\Theta }}[{c}(t)-{\theta }_{{\rm{p}}}]\nonumber\\ &-\frac{{\gamma }_{{\rm{d}}}\cdot\Delta t}{\tau_h}\cdot{h}(t)\cdot {{\Theta }}[{c}(t)-{\theta }_{{\rm{d}}}]. \nonumber
\end{align}
The early-phase weight dynamics allows us to divide the equation into three parts. The first two parts are related to plasticity in the presence of sufficient calcium while the third part always pull the early-phase weight to a steady-state value.
\begin{enumerate}
    \item Plasticity cases:
    \subitem Case of \gls{ltp}: $c\geq \theta_p > \theta_d$ yields
    \begin{equation}
        {h}(t+\Delta t) = h(t)\cdot(1-\frac{\Delta t}{\tau_h}\cdot({\gamma }_{{\rm{p}}}+{\gamma }_{\rm{d}})) +\frac{{\gamma }_{{\rm{p}}}\cdot\Delta t \cdot 255}{\tau_h},
    \end{equation}
    \subitem Case of \gls{ltd}: $\theta_p > c \geq \theta_d$ yields
    \begin{equation}
        {h}(t+\Delta t) = h(t)\cdot(1-\frac{{\gamma }_{\rm{d}}\cdot\Delta t}{\tau_h}).
    \end{equation}
    Since we multiply the early-phase weight with fractions $\dfrac{n}{d}$ for the two plasticity cases, we apply \gls{sr}, \cref{sr-round-eqn}, to the final value of \textit{h} as
    \begin{equation}
    \label{h-sr}
        h(t+\Delta t)=
        \begin{cases}
            h(t+\Delta t) + 1 \text{ if } P_h \leq \theta_h, \\
            h(t+\Delta t) \text{ if } P_h > \theta_h,
        \end{cases}
    \end{equation}
    where $\theta_h = \text{MAX\_UINT}_{32}\cdot\frac{(h(t)\cdot n)\bmod d}{d}$ and $P_h$ is a \SI{32}{\bit} random number.
    In the potentiation case, we have an additional fractional offset, so we also apply \gls{sr}, \cref{sr-update-eqn}, to this part.
	\item Steady-state: this part of the equation forces \textit{h} to converge to a steady-state value \textit{$h_0$} with a time constant of \textit{$10\cdot\tau_h$}
    \begin{equation}
        h(t+\Delta t)=
        \begin{cases}
            \text{$h(t)+1\cdot\mathrm{sgn}(h_0-h(t))$ if $P_{h, ss}\leq \theta_{h,ss}$},\\
            \text{$h(t)$ if $P_{h, ss}> \theta_{h,ss},$}
        \end{cases}
    \end{equation}
    with $\theta_{h,ss} = \text{MAX\_UINT}_{32}\cdot\dfrac{0.1\Delta t\cdot|h_0-h(t)|}{\tau_h}$ and $P_{h, ss}$ being a \SI{32}{\bit} random number. We chose $P_{h, ss}$ to be dependent on $h_0-h(t)$ so that $h(t)$ converges smoothly.
\end{enumerate}
The differential equation for the protein amount \textit{p} in \cref{protein-eqn} can be rewritten as
\begin{align}
    {p}(t+\Delta t)&={p}(t)-{p}(t)\cdot\frac{\Delta t}{\tau_p }\\&+\frac{\alpha \cdot\Delta t}{{\tau }_{p}}\cdot\ {{\Theta }}[\left|{h}(t)-{h}_{0}\right|-{\theta }_{{\rm{pro}}}] \nonumber.
\end{align}
Since $\frac{\Delta t}{{\tau }_{p}} < 1$ and $\frac{\alpha \cdot\Delta t}{{\tau }_{p}} < 1$, the evolution of the protein dynamics can be expressed as a sum comprising two summands:
\begin{enumerate}
    \item Case of protein synthesis during late-phase synaptic plasticity: $|h(t)-h_0|>\theta_{pro}$ yields
    \begin{equation}
		p(t + \Delta t)_{ps} =
        \begin{cases}
            p(t) + 1 \text{ if } P_{p, s} \leq \theta_{p,s}, \\
            p(t) \text{ if } P_{p, s} > \theta_{p,s},
        \end{cases}
    \end{equation}
    with $\theta_{p,s}=\text{MAX\_UINT}_{32}\cdot\dfrac{\alpha \cdot\Delta t}{{\tau }_{p}}$ and $P_{p,s}$ a \SI{32}{\bit} random number.
    \item Steady-state: this part of the equation allows the protein amount to decay back to \num{0} which yields
    \begin{equation}
		p(t + \Delta t)_{ss} =
        \begin{cases}
            p(t)-1 \text{ if } P_{p, ss} \leq \theta_{p, ss},\\
            p(t) \text{ if } P_{p, ss} > \theta_{p, ss},
        \end{cases}
    \end{equation}
    where $\theta_{p, ss}=\text{MAX\_UINT}_{32}\cdot{p}(t)\cdot\dfrac{\Delta t}{\tau_p }$ and $P_{p, ss}$ is a \SI{32}{\bit} random number. We chose $P_{p, ss}$ to be dependent on $p(t)$ so that $p(t)$ decays smoothly and more frequently.
\end{enumerate}
Finally, the differential equation for the late-phase weight \textit{z} in \cref{late-phase-eqn} can be rewritten as
\begin{align}
    {z}(t+\Delta t) &= {z}(t)\\ &+\frac{{p}(t)\cdot\Delta t}{255\cdot\tau_z}\cdot (127-{z}(t))\cdot {{\Theta }}[({h}(t)-{h}_{0})-{\theta }_{{\rm{tag}}}]\nonumber \\ &-\frac{{p}(t)\cdot\Delta t}{255\cdot\tau_z}\cdot ({z}(t)+64)\cdot {{\Theta }}[({h}_{0}-{h}(t))-{\theta }_{{\rm{tag}}}]. \nonumber
\end{align}
The evolution of the late-phase weight dynamics follows two cases:
\begin{enumerate}
    \item Case of late \gls{ltp}: ${h}(t)-{h}_{0}\geq{\theta }_{{\rm{tag}}}$ yields
    \begin{equation}
        z(t+\Delta t)_p=
        \begin{cases}
            z(t)+1 \text{ if } P_{z, p} \leq \theta_{z, p},\\
            z(t) \text{ if } P_{z, p} > \theta_{z, p},
        \end{cases}
    \end{equation}
    where $\theta_{z, p}=\text{MAX\_UINT}_{32}\cdot\dfrac{{p}(t)\cdot\Delta t}{255\cdot\tau_z}\cdot (127-{z}(t))$ and $P_{z, p}$ is a \SI{32}{\bit} random number.
    \item Case of late \gls{ltd}: ${h}_{0}-{h}(t)\geq{\theta }_{{\rm{tag}}}$ yields
    \begin{equation}
        z(t+\Delta t)_d=
        \begin{cases}
            z(t)-1 \text{ if } P_{z, d} \leq \theta_{z, d}, \\
            z(t) \text{ if } P_{z, d} > \theta_{z, d},
        \end{cases}
    \end{equation}
    where $\theta_{z, d}=\text{MAX\_UINT}_{32}\cdot\dfrac{{p}(t)\cdot\Delta t}{255\cdot\tau_z}\cdot ({z}(t)+64)$ and $P_{z, d}$ is a \SI{32}{\bit} random number.
\end{enumerate}


\section{Results}\label{sec:results}

\subsection{Emulating the Calcium Dynamics}

\begin{figure}
	\includegraphics[width=\linewidth]{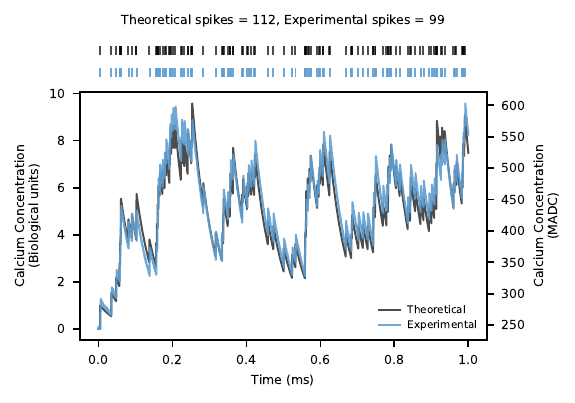}
	\caption{%
		Emulation of calcium dynamics recorded using the \gls{madc} at \SI{100}{kHz} stimulation frequency.
	   The theoretical calcium trace is calculated from Poisson spikes using \cref{calcium-eqn}. The experimental calcium trace is extracted using the adaptation trace of the \gls{adex} model from the recorded spikes. The theoretical spikes are Poisson spikes simulated at \SI{100}{kHz} during \SI{1}{ms}, and the experimental spikes are recorded from the parrot neuron.
	}
	\label{fig:calciumresults}
\end{figure}

The parameters of the adaptation signal are tuned to match the biological calcium trace (\cref{calcium-eqn}).
\Cref{fig:calciumresults} shows the results of the tuning for Poisson spikes simulated at a \SI{100}{kHz} stimulation frequency. 
Overall, the hardware-emulated calcium trace tracks the numerically computed calcium trace.
However, a slight misalignment exists for two reasons.
First, the adaptation circuit is non-linear beyond certain adaptation potentials.
Second, there is a probability of missing spikes. 
The misalignment from the nonlinearity of the circuit at high adaptation potentials can be reduced by further tuning specific parameters, specifically by reducing the spike-triggered adaptation \textit{b} parameter.
This reduces the adaptation potentials and allows the circuit to operate in the linear range at high frequency stimulations.
However, this comes with drawbacks at lower stimulation frequencies where the recorded adaptation potentials are not high, and where the depression threshold $\theta_d^{hw}=b$ can be easily corrupted by noise.

\subsection{Running the Plasticity Algorithm}

\begin{figure}
	\includegraphics[width=\linewidth]{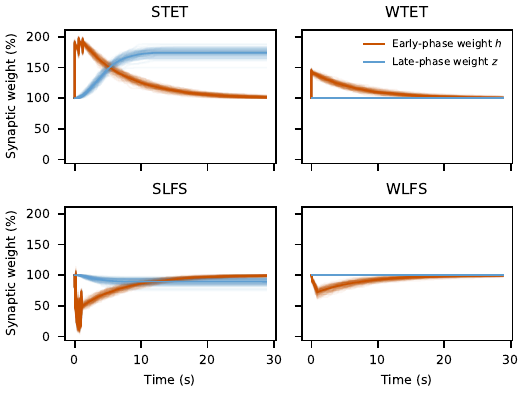}
	\caption{
		Emulation results for the four stimulation protocols for one spike trial and \num{100} different update seeds.
		The trajectories in faded color show the individual trials, and the trajectory in bold shows the average behavior.
		The overall average behavior is in agreement with the behavior obtained by simulation for the four protocols.
	}
	\label{fig:resultssamespikes}
\end{figure}

\begin{figure}
	\includegraphics[width=\linewidth]{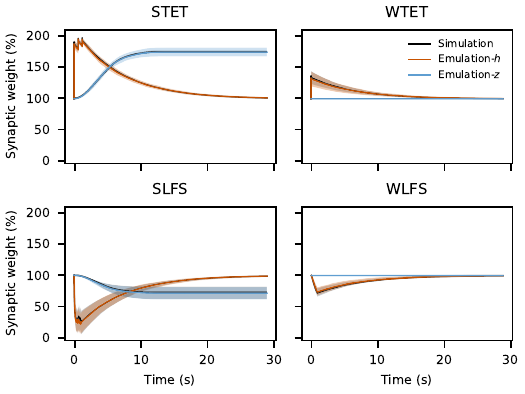}
	\caption{
		Emulation results for the four stimulation protocols compared against a simulation baseline at a time step of \SI{50}{ms} for \num{100} different spike trials.
		The lines correspond to the average early-phase and late-phase weights.
		The bands correspond to one standard deviation from the average.
		For comparison purposes, the simulation results are plotted at an accelerated factor of \num{1000} to match the time of the emulation results.
	}
	\label{fig:resultsdiffspikes}
\end{figure}

The differential equations are solved with a numerical timestep $\Delta t = \SI{50}{\micro\second}$.
First, we run the experiment for \num{100} different update seeds and a fixed input spike pattern.
Due to updating variables at different iterations, each individual update seed yields a different trace but a common overall behavior, \cref{fig:resultssamespikes}.
Averaging across the update seeds produces a behavior similar to the one obtained from a simulation at a timestep of \SI{50}{ms}, compare \cref{fig:resultsdiffspikes}.

Second, we run the experiment for \num{100} spike trials; as in \cite{luboeinski2021memory}, the target is to obtain the average behavior of the synaptic weights to the stimulation protocols across different spike trials.
We plot the results against the simulation baseline at a timestep of $\SI{50}{\milli\second}_\text{bio}$ in figure \ref{fig:resultsdiffspikes}. For the simulation baseline, we also use the concept of multi-timestep updates where the membrane potential and calcium concentration are updated at a timestep $\SI{0.2}{\milli\second}_\text{bio}$ while synaptic weights and protein amount are updated at $\SI{50}{\milli\second}_\text{bio}$.

The mean of the synaptic weights aligns with those obtained from the simulation with small differences.
The variances of the emulation are close to the simulation for all protocols except for the \gls{stet} protocol.
This can be attributed to the robustness of the \gls{stet} protocol and the dominance of the effects that arise from the analog nature of the hardware and stochastic rounding.
For the other protocols, the results show that the effects in the hardware variability seem to cancel out with different spike trials.

\section{Discussion}\label{sec:discussion}

We implemented and validated a calcium-based plasticity rule integrating the \gls{stc} hypothesis on \gls{bss2}.
Our approach faithfully reproduces the simulation results of the single synapse for four typical stimulation protocols with small differences in the mean and standard deviation.
This highlights the power of stochastic rounding for integer arithmetic and reduced precision.
The differences between the emulation at a time update of $\SI{50}{\micro\second}_\text{hw} = \SI{50}{\milli\second}_\text{bio}$ and the simulation at a time update of $\SI{50}{\milli\second}$ can be primarily attributed to the variability in the calcium dynamics due to the analog nature of the hardware.
The weight update is sensitive to this variability because the calcium dynamics are fast and can decay quickly or rise almost instantaneously by the effect of spikes.

Another important consideration is tuning the parameters of the adaptation circuit that is used to emulate the calcium dynamics.
The tuning has to account for circuit linearity and the operation at all considered stimulation frequencies.
This presents a challenge since there is a trade-off between the tuning at low frequencies and high frequencies.
To further improve the results, we may optimize the tuning process of the adaptation parameters using numerical tests and an iterative process.

We consider our implementation of the plasticity rule successful since the overall variability stems from the spike variability, as in the simulation case.
In other words, the effects that arise from the analog nature of the hardware and stochastic rounding are compensated across different trials.
This is clear from the experiments across different spike trials, where the mean and variance of the synaptic weights in the emulation are almost equal to those of the simulation case in the four stimulation protocols.
For the \gls{stet} protocol, these effects are more apparent, as indicated by the high variance compared to the simulation.
The reason is that the \gls{stet} protocol is considered robust due to its high calcium concentration, making it less prone to the variability in calcium dynamics.

Our work can be expanded to emulate networks on \gls{bss2} with neurons that follow the investigated plasticity rule, similar to \cite{luboeinski2021memory}.
The network emulation on \gls{bss2} is made possible through built-in units that undergo parallel computations \cite{billaudelle2021structural, cramer2022surrogate}.
We expect that the reduced runtime of the network emulation due to the acceleration factor of \gls{bss2} allows to run experiments for different network topologies, neuron parametrization, and a higher number of spike trials.

This study showcases that analog neuromorphic hardware can facilitate the investigation of research questions that are unsuitable for numerical simulations due to long simulation runtimes.
In principle, accelerated neuromorphic substrates are well-suited for extended-duration experiments, parameter sweeps, or iterative experiment reconfiguration.
However, the use of neuromorphic systems comes at a cost:
network topologies, model parametrization, and dynamic behavior of neurons and synapses, as well as plasticity rules must be adapted to the substrate.
Especially for neuroscientific research questions, we expect neuromorphic hardware to be complemented by traditional simulation methods.
In this complementary approach, the exploration of experiment parametrization, and, in particular, the directed evaluation and reconfiguration of experiments should be offloaded to neuromorphic accelerators,
while the simulator provides validation, digital reproducibility, and high numerical precision.

This purpose can be served well by \Gls{bss2}, as its integration into the open research infrastructure \mbox{EBRAINS}\footnote{\url{https://ebrains.eu}} has enabled the global research community to access the neuromorphic platform.
\todo{%
ECM4AA: Runtime performance; and it would be really nice if we could refer to some tutorial at some point --- e.g., for the journal pub :).
}

\section*{Acknowledgements}
\addcontentsline{toc}{section}{Acknowledgment}

This work has received funding from
the EC Horizon Europe Framework Programme
under grant agreement
101147319 (EBRAINS 2.0),
the \foreignlanguage{ngerman}{Deutsche Forschungsgemeinschaft} (DFG, German Research Foundation) under Germany's Excellence Strategy EX 2181/1-390900948 (the Heidelberg STRUCTURES Excellence Cluster) as well as through grants SFB1286/C01\&Z01, TE 1172/7-1 [CT] and the \foreignlanguage{ngerman}{Bundesministerium für Bildung und Forschung (BMBF)}, grant number 01IS22093A-E [CT].

\section*{Author Contributions}\label{sec:author_contributions}

We give contributions in the \textit{CRediT} (Contributor Roles Taxonomy) format:
\textbf{AA}: Investigation, visualization, methodology, software;
\textbf{JK}: Conceptualization, methodology, supervision, software, visualization;
\textbf{SB}: Conceptualization, methodology, supervision;
\textbf{PS}: Validation, software, resources;
\textbf{EM}: Conceptualization, methodology, supervision, software, resources;
\textbf{JL}: Validation, methodology, software;
\textbf{CT}: Validation, methodology, funding acquisition;
\textbf{JS}: Conceptualization, methodology, supervision, funding acquisition;
\textbf{all}: writing --- original draft, writing --- reviewing \& editing.


\printbibliography

@ARTICLE{brette2005adaptive,
  author = {Brette, R. and Gerstner, W.},
  title = {Adaptive Exponential Integrate-and-Fire Model as an Effective Description
	of Neuronal Activity},
  journal = {J. Neurophysiol.},
  year = 2005,
  volume = {94},
  pages = {3637 -- 3642},
  affiliation = {EPFL},
  details = {http://infoscience.epfl.ch/record/97829},
  documenturl = {http://infoscience.epfl.ch/getfile.py?mode=best&recid=97829},
  doi = {10.1152/jn.00686.2005},
  oai-id = {oai:infoscience.epfl.ch:97829},
  oai-set = {article; fulltext},
  review = {REVIEWED},
  status = {PUBLISHED},
  unit = {LCN},
}

@article{graupner2012calcium,
  title={Calcium-based plasticity model explains sensitivity of synaptic changes to spike pattern, rate, and dendritic location},
  author={Graupner, Michael and Brunel, Nicolas},
  journal={Proc. Natl. Acad. Sci.},
  volume=109,
  number=10,
  pages={3991--3996},
  year=2012,
  publisher={National Academy of Sciences},
  doi={10.1073/pnas.1109359109}
}

@article{dietz2015understanding,
  title={Understanding integer overflow in {C/C++}},
  author={Dietz, Will and Li, Peng and Regehr, John and Adve, Vikram},
  journal={ACM Trans. Softw. Eng. Methodol.},
  volume=25,
  number=1,
  pages={1--29},
  issn = {1049-331X},
  year=2015,
  month=dec,
  publisher={Association for Computing Machinery},
  address={New York, NY, USA},
  doi={10.1145/2743019},
}

@article{forsythe1950round,
  title={Round-off errors in numerical integration on automatic machinery-preliminary report},
  author={Forsythe, George E},
  journal={Bull. Amer. Math. Soc},
  volume=56,
  pages={61--62},
  year=1950
}

@article{frey1997synaptic,
  title={Synaptic tagging and long-term potentiation},
  author={Frey, Uwe and Morris, Richard G. M.},
  journal={Nature},
  volume=385,
  number=6616,
  pages={533--536},
  year=1997,
  publisher={Nature Publishing Group},
  doi={10.1038/385533a0}
}

@INPROCEEDINGS{hock13analogmemory,
	author = {Hock, M. and Hartel, A. and Schemmel, J. and Meier, K.},
	booktitle = {Circuit Theory and Design (ECCTD), 2013 European Conference on},
	title = {An analog dynamic memory array for neuromorphic hardware},
	year = 2013,
	pages = {1-4},
	doi = {10.1109/ECCTD.2013.6662229},
	month = sep,
}

@article{hopkins2020stochastic,
  title={Stochastic rounding and reduced-precision fixed-point arithmetic for solving neural ordinary differential equations},
  author={Hopkins, Michael and Mikaitis, Mantas and Lester, Dave R. and Furber, Steve},
  journal={Phil. Trans. R. Soc. A},
  volume=378,
  number=2166,
  pages=20190052,
  year=2020,
  publisher={The Royal Society Publishing},
  doi={10.1098/rsta.2019.0052}
}

@misc{kahan1996ieee,
  title={{IEEE} standard 754 for binary floating-point arithmetic},
  author={Kahan, William M.},
  note={Lecture Notes on the Status of IEEE},
  year=1996,
  url={https://web.archive.org/web/20020622093102/http://www.cs.berkeley.edu/~wkahan/ieee754status/IEEE754.PDF}
}

@article{luboeinski2021memory,
  title={Memory consolidation and improvement by synaptic tagging and capture in recurrent neural networks},
  author={Luboeinski, Jannik and Tetzlaff, Christian},
  journal={Commun. Biol.},
  volume=4,
  number=1,
  pages=275,
  year=2021,
  month = mar,
  pages = {1--17},
  issn = {2399-3642},
  publisher={Nature Portfolio},
  doi={10.1038/s42003-021-01778-y}
}

@article{marsaglia2003xorshift,
  title={Xorshift {RNG}s},
  author={Marsaglia, George},
  journal={Journal of Statistical software},
  volume=8,
  number=14,
  pages={1--6},
  year=2003,
  doi={10.18637/jss.v008.i14}
}

@ARTICLE{mueller2022scalable_noeprint,
  author = {M{\"u}ller, Eric and Arnold, Elias and Breitwieser, Oliver and Czierlinski, Milena and Emmel, Arne and Kaiser, Jakob and Mauch, Christian and Schmitt, Sebastian and Spilger, Philipp and Stock, Raphael and Stradmann, Yannik and Weis, Johannes and Baumbach, Andreas and Billaudelle, Sebastian and Cramer, Benjamin and Ebert, Falk and Göltz, Julian and Ilmberger, Joscha and Karasenko, Vitali and Kleider, Mitja and Leibfried, Aron and Pehle, Christian and Johannes Schemmel},
  title = {A Scalable Approach to Modeling on Accelerated Neuromorphic Hardware},
  volume = 16,
  year = 2022,
  journal = {Front. Neurosci.},
  doi = {10.3389/fnins.2022.884128},
  issn = {1662-453X},
}

@TECHREPORT{powerisa_206,
  author = {{PowerISA}},
  title = {Power{ISA} Version 2.06 Revision B},
  institution = {Power.org},
  year = 2010,
  month = jul,
  url = {http://www.power.org/resources/reading/},
  type = {Specification},
}

@article{redondo2011making,
  title={Making memories last: the synaptic tagging and capture hypothesis},
  author={Redondo, Roger L. and Morris, Richard G. M.},
  journal={Nat Rev Neurosci},
  issn={1471-0048},
  volume=12,
  number=1,
  pages={17--30},
  year=2011,
  publisher={Nature Publishing Group},
  doi={10.1038/nrn2963}
}

@inproceedings{wang2013towards,
  author = {Wang, Xi and Zeldovich, Nickolai and Kaashoek, M. Frans and Solar-Lezama, Armando},
  title = {Towards optimization-safe systems: analyzing the impact of undefined behavior},
  year = 2013,
  isbn = {9781450323888},
  publisher = {Association for Computing Machinery},
  address = {New York, NY, USA},
  doi = {10.1145/2517349.2522728},
  booktitle = {Proceedings of the Twenty-Fourth ACM Symposium on Operating Systems Principles},
  pages = {260--275},
  numpages = 16,
  location = {Farminton, Pennsylvania},
  series = {SOSP '13}
}

@ARTICLE{zenke2014auryn, crossref = {zenke2014limits}}

@ARTICLE{zenke2014limits,
	AUTHOR={Zenke, Friedemann  and  Gerstner, Wulfram},
	TITLE={Limits to high-speed simulations of spiking neural networks using general-purpose computers},
	JOURNAL={Front. Neuroinform.},
	VOLUME=8,
	YEAR=2014,
	NUMBER=76,
	URL={http://www.frontiersin.org/neuroinformatics/10.3389/fninf.2014.00076/abstract},
	DOI={10.3389/fninf.2014.00076},
	ISSN={1662-5196},
}

@article{cramer2022surrogate,
  title={Surrogate gradients for analog neuromorphic computing},
  author={Cramer, Benjamin and Billaudelle, Sebastian and Kanya, Simeon and Leibfried, Aron and Gr{\"u}bl, Andreas and Karasenko, Vitali and Pehle, Christian and Schreiber, Korbinian and Stradmann, Yannik and Weis, Johannes and others},
  journal={Proc. Natl. Acad. Sci.},
  volume=119,
  number=4,
  year=2022,
  publisher={National Acad Sciences},
  doi={10.1073/pnas.2109194119},
}

@article{croci2022stochastic,
  title   = {Stochastic rounding: Implementation, error analysis and applications},
  author  = {Croci, Matteo and Fasi, Massimiliano and Higham, Nicholas J. and Mary, Theo and Mikaitis, Mantas},
  journal   = {R. Soc. Open Sci.},
  volume    = 9,
  number    = 3,
  pages     = 211631,
  year      = 2022,
  publisher = {The Royal Society Publishing},
  doi       = {10.1098/rsos.211631}
}

@Article{billaudelle2021structural,
  author          = {Billaudelle, Sebastian and Cramer, Benjamin and Petrovici, Mihai A. and Schreiber, Korbinian and Kappel, David and Schemmel, Johannes and Meier, Karlheinz},
  journal         = {Neural networks: the official journal of the International Neural Network Society},
  title           = {Structural plasticity on an accelerated analog neuromorphic hardware system},
  year            = 2021,
  issn            = {1879-2782},
  month           = jan,
  pages           = {11--20},
  volume          = {133},
  doi             = {10.1016/j.neunet.2020.09.024},
}

@article{pehle2022brainscales2_nopreprint_nourl,
  author  = {Christian Pehle and Sebastian Billaudelle and Benjamin Cramer and Jakob Kaiser and Korbinian Schreiber and Yannik Stradmann and Johannes Weis and Aron Leibfried and Eric M{\"u}ller and Johannes Schemmel},
  title   = {The {BrainScaleS-2} Accelerated Neuromorphic System with Hybrid Plasticity},
  journal = {Front. Neurosci.},
  volume  = 16,
  year    = 2022,
  doi     = {10.3389/fnins.2022.795876},
  issn    = {1662-453X},
}

@INPROCEEDINGS{billaudelle2022accurate,
  author = {Billaudelle, Sebastian and Weis, Johannes and Dauer, Philipp and Schemmel, Johannes},
  booktitle = {ICECS},
  title = {An accurate and flexible analog emulation of {AdEx} neuron dynamics in silicon},
  year = 2022,
  pages = {1--4},
  doi = {10.1109/ICECS202256217.2022.9971058}
}

@article{sajikumar2005synaptic,
  title={Synaptic tagging and cross-tagging: the role of protein kinase M$\zeta$ in maintaining long-term potentiation but not long-term depression},
  author={Sajikumar, Sreedharan and Navakkode, Sheeja and Sacktor, Todd Charlton and Frey, Julietta Uta},
  journal={Journal of Neuroscience},
  volume={25},
  number={24},
  pages={5750--5756},
  year={2005},
  publisher={Soc Neuroscience},
  doi={10.1523/JNEUROSCI.1104-05.2005}
}

@incollection{luboeinski2024modeling,
  title={Modeling Emergent Dynamics Arising from Synaptic Tagging and Capture at the Network Level},
  author={Luboeinski, Jannik and Tetzlaff, Christian},
  booktitle={Synaptic Tagging and Capture: From Synapses to Behavior},
  pages={471--503},
  year={2024},  
  edition   = {2nd},
  address   = {Cham, Switzerland},
  chapter   = {23},
  editor    = {Sajikumar, Sreedharan and Abel, Ted},
  publisher = {Springer},
  doi       = {10.1007/978-3-031-54864-2_23}
}

@incollection{prodan2024synaptic,
  title={Synaptic Tagging and Capture: Functional Implications and Molecular Mechanisms},
  author={Prodan, Alex and Morris, Richard GM},
  booktitle={Synaptic Tagging and Capture: From Synapses to Behavior},
  pages={1--41},
  year={2024},
  edition   = {2nd},
  address   = {Cham, Switzerland},
  chapter   = {1},
  editor    = {Sajikumar, Sreedharan and Abel, Ted},
  publisher={Springer},
  doi={10.1007/978-3-031-54864-2_1}
}

@article{li2016induction,
  title={Induction and consolidation of calcium-based homo-and heterosynaptic potentiation and depression},
  author={Li, Yinyun and Kulvicius, Tomas and Tetzlaff, Christian},
  journal={PlOS one},
  volume={11},
  number={8},
  pages={e0161679},
  year={2016},
  publisher={Public Library of Science San Francisco, CA USA},
  doi={10.1371/journal.pone.0161679}
}

@ARTICLE{okuda2021initial,
  author    = {Okuda, Kosuke and H{\o}jgaard, Kristoffer and Privitera, Lucia and Bayraktar, G{\"u}lberk and Takeuchi, Tomonori},
  journal   = {European Journal of Neuroscience},
  title     = {Initial memory consolidation and the synaptic tagging and capture hypothesis},
  year      = 2021,
  pages     = {6826--6849},
  volume    = 54,
  number    = 8,
  doi       = {10.1111/ejn.14902}
}

@Article{shires2012synaptic,
    author  = {Shires, K. L. and Da Silva, B. M. and Hawthorne, J. P. and Morris, R. G. M. and Martin, S. J.},
    title   = {Synaptic tagging and capture in the living rat},
    journal = {Nature Communications},
    year    = 2012,
    volume  = 3,
    number  = 1,
    pages   = {1246},
    issn    = {2041-1723},
    doi     = {10.1038/ncomms2250},
}

@Article{moncada2007induction,
    author = {Moncada, Diego and Viola, Hayd{\'e}e},
    title = {Induction of Long-Term Memory by Exposure to Novelty Requires Protein Synthesis: Evidence for a Behavioral Tagging},
    volume = 27,
    number = 28,
    pages = {7476--7481},
    year = 2007,
    doi = {10.1523/JNEUROSCI.1083-07.2007},
    issn = {0270-6474},
    journal = {Journal of Neuroscience}
}

@Article{moncada2015behavioral,
  title    = {Behavioral Tagging: A Translation of the Synaptic Tagging and
              Capture Hypothesis},
  author   = {Moncada, Diego and Ballarini, Fabricio and Viola, Hayd{\'e}e},
  journal  = {Neural Plast},
  volume   = 2015,
  pages    = {650780},
  year     = 2015,
  doi      = {10.1155/2015/650780}
}

@article{ziegler2015synaptic,
    author = {Ziegler, Lorric and Zenke, Friedemann and Kastner, David B. and Gerstner, Wulfram},
    title = {Synaptic Consolidation: From Synapses to Behavioral Modeling},
    volume = 35,
    number = 3,
    pages = {1319--1334},
    year = 2015,
    doi = {10.1523/JNEUROSCI.3989-14.2015},
    issn = {0270-6474},
    journal = {Journal of Neuroscience}
}

@article{clopath2008tag,
    title   = {Tag-trigger-consolidation: A model of early and late long-term-potentiation and depression},
    volume  = 4,
    doi     = {10.1371/journal.pcbi.1000248},
    number  = 12,
    journal = {PLoS Computational Biology},
    author  = {Clopath, Claudia and Ziegler, Lorric and Vasilaki, Eleni and Büsing, Lars and Gerstner, Wulfram},
    year    = 2008}

\end{document}